%% file: CDGCN.tex
\documentclass[conference,a4paper]{IEEEtran}

\usepackage{CJKutf8}
\usepackage[switch]{lineno}
\usepackage{makecell}
\usepackage{graphicx}
\usepackage{upgreek}
\usepackage{CJK}
\usepackage{bbm}
\usepackage{hyperref} 
\usepackage{graphicx}
\usepackage{float}
\usepackage{lettrine}
\usepackage{booktabs}
\usepackage{bm}
\usepackage{amsmath}
\DeclareRobustCommand*{\IEEEauthorrefmark}[1]{\raisebox{0pt}[0pt][0pt]{\textsuperscript{\footnotesize #1}}}

\usepackage{fontawesome}
\usepackage[table,xcdraw]{xcolor}
\usepackage{soul}
\begin{document}
\IEEEoverridecommandlockouts
\pagenumbering{roman}

%
\title{Multi-Scale U-Shape MLP for Hyperspectral Image Classification}

\definecolor{Moule}{rgb}{1.0,0.1,0.1}
\newcommand{\Moule}[1]{{\textbf{\color{Moule} #1}}}

\newcommand{\REV}[1]{{\textbf{\color{Rev1st} #1}}}
\newcommand{\ourM}{{MUMLP}}
\def\eg{\emph{e.g.}} \def\Eg{\emph{E.g}}
\def\ie{\emph{i.e.}} \def\Ie{\emph{I.e.}}
\def\cf{\emph{c.f.}} \def\Cf{\emph{C.f.}}
\def\etc{\emph{etc}} \def\vs{\emph{vs.}}
\def\aka{\emph{a.k.a.}}
\def\wrt{w.r.t.} \def\dof{d.o.f.}
\def\etal{\emph{et al.}}
\author{\IEEEauthorblockN{
Moule Lin$^{\href{https://orcid.org/0000-0001-6227-2392}{\includegraphics[scale=0.07]{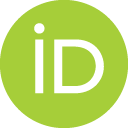}}}$,\thanks{The work described in this paper is supported by National Natural Science Foundation of China (31770768), Fundamental Research Funds for the Central Universities(2572017PZ04), Heilongjiang Province Applied Technology Research and Development Program Major Project(GA18B301,GA20A301) and China State Forestry Administration Forestry Industry Public Welfare Project (201504307) }\thanks{(Corresponding author: Weipeng Jing.)}   
Weipeng Jing$^{\href{https://orcid.org/0000-0001-7933-6946}{\includegraphics[scale=0.07]{images/ID.png}}}$, \textit{Member, IEEE},\thanks{M. Lin, W. Jing and G. Chen are with the College of Information and Computer Engineering, Northeast Forestry University, Harbin 150040, China.  (e-mail: nefu\_ml@nefu.edu.cn, jwp@nefu.edu.cn, kjc\_chen@163.com)}   
Donglin Di$^{\href{https://orcid.org/0000-0002-2270-3378}{\includegraphics[scale=0.07]{images/ID.png}}}$,\thanks{D. Di is with the Baidu Co., Ltd, Beijing 100085, China (e-mail: didonglin@baidu.com).}  
Guangsheng Chen, \textit{ Member, IEEE}\\
and Houbing Song$^{\href{https://orcid.org/0000-0003-2631-9223}{\includegraphics[scale=0.07]{images/ID.png}}}$, \textit{Senior Member, IEEE}\thanks{H. Song is with the Department of Electrical, Computer, Software, and Systems Engineering, Embry-Riddle Aeronautical University, Daytona Beach, FL 32114, USA (e-mail: SONGH4@erau.edu)}
}    
}
\maketitle
\input{section/0-abstract}

\input{section/1-introduction}

\input{section/2-propose}

\input{section/3-experiments}

\input{section/4-conclusion}

\input{section/5-Acknowledgement}
\IEEEpeerreviewmaketitle
\vspace{7pt}
\bibliographystyle{IEEEtran}
\bibliography{REFERENCE}
\end{document}

%% file: section/0-abstract.tex
\begin{abstract}
Hyperspectral images have significant applications in various domains, since they register numerous semantic and spatial information in the spectral band with spatial variability of spectral signatures.
Two critical challenges in identifying pixels of the hyperspectral image are respectively representing the correlated information among the local and global, as well as the abundant parameters of the model.
To tackle this challenge, we propose a Multi-Scale U-shape Multi-Layer Perceptron (\ourM) a model consisting of the designed MSC (Multi-Scale Channel) block and the UMLP (U-shape Multi-Layer Perceptron) structure.
MSC transforms the channel dimension and mixes spectral band feature to embed the deep-level representation adequately.
UMLP is designed by the encoder-decoder structure with multi-layer perceptron layers, which is capable of compressing the large-scale parameters.
Extensive experiments are conducted to demonstrate our model can outperform state-of-the-art methods across-the-board on three wide-adopted public datasets, namely Pavia University, Houston 2013 and Houston 2018. 
\end{abstract}
{\IEEEkeywords Hyperspectral Image (HSI), Multi-Scale Shape, Multi-Layer Perceptron, Compression Model}

%% file: section/1-introduction.tex
\section{INTRODUCTION}
\lettrine[lines=2]{H}{YPERSPECTRAL} Image (HSI) contains complex multiple structures and composed by numerous bands with location and distribution information \cite{luo2019sparse}.
Therefore, it has a significant value for monitoring the Earth’s surface, and the research of hyperspectral remote sensing images has broad applications in society and other domains. 
An incomplete list includes environmental sciences \cite{pandey2020new}, agronomy \cite{ahangarha2020crop}, geography \cite{song2017deep}, astronomy \cite{Rodet2009DataIF}, mineralogy \cite{Fox2016ApplicationsOH}, and so on.
The hyperspectral remote sensing images research mainly focuses on analyzing the spectral band's complex intra-class and inter-class relationships.\par
The scientific research of hyperspectral has a long history, from artificial geometry to deep learning, all of which have extensively promoted the progress of hard science and society.
A.Huertas \etal \, \cite{huertas1988detecting} was the first one to apply hyperspectral image in urban analysis. 
\begin{figure}
    \centering
    \includegraphics[width=0.5\textwidth]{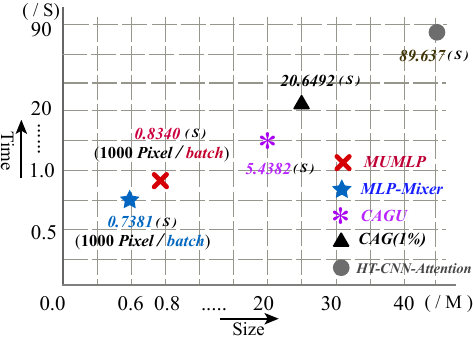}
    \vspace{-0.8cm}
    \caption{Model size and consumption of computational for five methods. MsUMLP and MLP-Mixer adopted 1000 pixels per batch as a benchmark.}
    \vspace{-1cm}
    \label{fig:0}
\end{figure}
It adopted geometric topology theory to extract the urban building boundary. JB Adams \etal \, \cite{adams1989simple} analyzes the spectral band through mathematical theory and combine other factors to analyze hyperspectral data. 
In general, early hyperspectral image research primarily employed theoretical knowledge of various subjects, such as geometry, to analyze the shallow information.
The state-of-the-art methods always allude to deep learning represented by convolutional neural networks (CNNs).
Moreover, it also got a colossal achievement and outstanding performance in semantic segmentation, classification, target detection, and so forth.
Luo \etal\, proposed semisupervised sparse manifold discriminative analysis ((S$^3$MDA)\cite{luo2016semisupervised} to overcome the challenge of how to graph embedding (GE) selecting a proper neighborhood size for graph construction. Furthermore, they proposed a more advanced method called enhanced hybrid-graph discriminant learning (EHGDL) structure that adopted intraclass hypergraph and an interclass hypergraph to analyze the complex multiple relationships of an HSI. \cite{luo2020dimensionality}
The role of CNNs in deep learning is undeniably likely a de-facto standard. 
%
Its parameter, however, increases exponentially with the layers of convolution, and its size increases with computational capacity, consistently exceeding 20M to have a qualified express ability.
Moreover, computational consumption is a bottleneck for industrial applications due to the long-lasting operation of multiplication and addition, which cannot satisfy the real-time requirement of the industry.
\par
%
%

%
I. Tolstikhin proves that CNNs are not necessary for deep learning. The present MLP-Mixer \cite{Tolstikhin2021MLPMixerAA} frame adopted two types of MLPs to mix per-location features and spatial information, respectively. It is a meaningful research topic. 
By sacrificing the microscopic accuracy, the model speed has improved remarkably, and the model size is also compressed.
Despite its simplicity, it has outstanding performance in various domains and disciplines. And many kinds of research are based on MLP-Mixer in the remote sensing domain to excavate more meaningful analysis. 
%
%
M. Lin \etal \, \cite{Lin2021ContextAwareAG} \, employed GCN to transform the features from a chaotic state into a high cohesion state with reducing the data’s redundant information. 
And the noise in the hyperspectral images is also a challenging problem. 
%
HyMiNoR \cite{Rasti2020HyperspectralMG} is also an effective denoising method using a novel sparse noise frame.
In addition, U-Net \cite{Ronneberger2015UNetCN} is a classic encoder-decoder structure in which the encoder embeds spatial and semantic information and the decoder mix that information with position features. 
\par
Although existing methods can perform well in hyperspectral classification, their model's computational consumption is terrible, and its running time is always long-winded.
The convolution operation is the source of all problems. It brings outstanding performance in various fields, but it also brings a complicated calculation consumption. 
MLP-Mixer is epoch-making research that only contains MLP operation by stacking layers to mix all features, such as spatial information.
However, its expressive ability is weakened due to its simple model structure, mainly by ignoring the semantic information between the neighbouring structures. 
\par
To address the problems mentioned above, We propose \ourM~(Multi-scale U-shape's Multi-Layer Perceptron) that the model's parameter scale is only 0.817M, and the running consumption fully satisfies the demand of the industry. The main contributions of this letter are summarized as follows:
\par

\begin{itemize}
\item[1)]
According to the idea of our model. \ourM~is constituted of MSC (Multi-scale channel) block and UMLP (U-shape's Multi-Layer Perceptron) block.
And the main contributions of our work in this letter is that its parameter scale is only 0.817M. What's more, to enhance each pixels feature expression, we proposed the MSC block that transforms the channel dimension to ${{2}^{n}}$ (\rm{n} is an integer) to unify the channel data of different datasets. The 1$\times$1 convolution operation can easily excavate deeper, was adopted to expands the representation of various features of the pixel to obtain the distribution of multiple dimensions. stacking MSC-2 to get deeper receptive fields of module employed for Gen-C channels is a tack strategy for features extracted of entirety model, which solves the bottleneck of crucial information loss without causing parameter explosion.
\par
\end{itemize}
\begin{itemize}
\item[2)]
In addition, \ourM~solves the drawback of MLP-Mixer's poor expressive ability according to an encoder-decoder structure to effectively increase the model's embedding ability for deep semantic information. Multi-Layer perceptron is a simple, but efficient when variant features contain in the previous phase to next, which does not involve complex multiplication, and its size is only 0.817M. Unlikely normal U-Net, each step existed one process only, this component designate three different steps, including one rotation and two MLP operations. The skip connection combines deeply semantic information and spatial information. The U-block will improve the discriminant performance of different features types and for each class. Compare the state-of-the-art techniques as shown in Fig.~\ref{fig:0}.
\end{itemize}

\begin{figure*}
    \centering
    \includegraphics[width=1.0\textwidth]{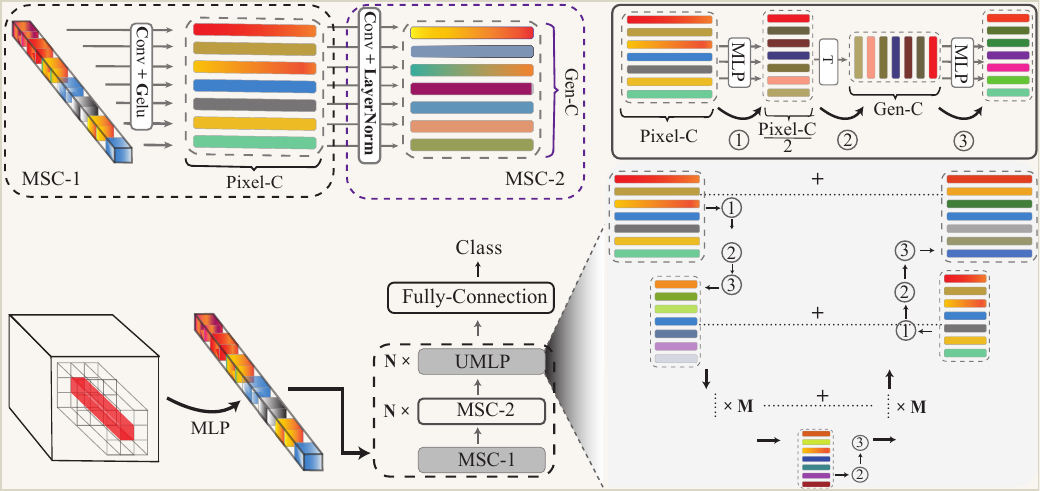}
    \vspace{-0.5cm}
    \caption{\ourM~Architecture. The bottom half of the figure is the main component model, from input to class. Which mainly includes MSC module and UMLP module. MSC module: It is composed of MSC1 and MSC2 shown on the upper right. UMLP module: the right side of the bottom half of the figure. Both MSC and UMLP have to be stacked N times to get a deeper receptive field.}
    \vspace{-0.52cm}
    \label{fig:my_label}
\end{figure*}

%% file: section/2-propose.tex
\section{propose}
We transformed the segmentation into a classification task in the proposed method, working on pixel-wise classification instead of patch segmentation. 
%
%
It outputs each pixel of hyperspectral will form an image-like patch. Each row represents a different feature expression of each pixel generated by convolution, hereinafter reference as: Pixel-$C$. 
Each column represents the generalization of the original pixel channel value, hereinafter reference as: Gen-$C$.\par
MSC mainly contains two parts: one with MSC applied to extract the features of Pixel-$C$, similar with pooling operation, and one with MSC employed to mix Gen-$C$ information.\par
For hyperspectral image, its size is ${{\mathbf{\tilde{I}}}^{\mathbf{H} \times \mathbf{W} \times \mathbf{C}}}$. \textit{C} is the number of spectral bands, H and W are the image height and width respectively.
The random method shuffled all the pixels and excluded background, adopted to sample each pixel randomly. It is described as follows:
\begin{equation}
    {{\mathbf{I}}^{\mathbf{B}\times 1\times \mathbf{C}}}={{\mathbf{S}}_{\mathbf{k}}}\left( {{{\tilde{\mathbf{I}}}}^{\mathbf{B}\times \mathbf{H}\times \mathbf{W}\times \mathbf{C}}} \right)
\end{equation}
$S$ is a random function, and B represents the batch size. $k$ denotes the sampling range, which determines whether it is the training or test set. On the whole, the input data is a pixel point rather than a patch.
\par
In the MSC block, we employed two MLP layers to extract the deeper semantic information to express the pixel features about its kernel attributes, which can be written as follows:
\begin{equation}
    {{\mathbf{X}}_{*,i}}=\mathbf{\Gamma} \left( {{\mathbf{W}}_{2}}\left( \sigma \left( {{\mathbf{W}}_{1}}\mathit{\tau}\left( {{\mathbf{I}}_{*,i}} \right) \right) \right) \right),\text{    }\mathit{for}\text{ }i=1\ldots \mathit{c}
\end{equation}
Here $c$ is Pixel-$C$ and $\tau$ is LayerNorm. $\mathbf{{W}}_{1}$ and $\mathbf{{W}}_{2}$ are weight of tow MLP layers respectively. $\sigma$ represents a nonlinearity (GELU). $\mathbf{\Gamma}$ denotes dropout function.\par
The Pixel-$C$ dimension only embeds one feature of pixels. We employ, therefore, convolution operation with $1\times1$ size of the kernel, which extracts more semantic features in which embedded various information to generalize each pixels channel and named Gen-$C$. It describes as the following formula:
\begin{equation}
    {{\mathbf{U}}_{*,j,*}}=\tau\left( Conv\left( {{\mathbf{X}}_{*,j,*}} \right) \right),\text{    }\mathit{for}\text{ }j=1\ldots \mathit{cls}
\end{equation}
The Gen-$C$ of pixels features $ {{\mathbf{U}}_{*,j,*}}$ relies on the number of pixel classes $\mathit{cls}$. The size of the convolution kernel is $1 \times 1$. 
We adopted a convolutional layer with a small parameter to generalize each pixel into a deeper semantic level.\par
For UMLP block, it is composed of MixerBlock, U-shape interpret and skip connection modules, respectively. 
Stacking the MSC layer to get a higher receptive field, the input $ ({{\mathbf{U}}_{*,j,*}})$ has global feature information on both the Pixel-$C$ dimension and Gen-$C$. 
The MixerBlock module mixes semantic information in two directions through two MLP layers and then extracts Pixel-C dimension features through one MLP (reduced dimension), as shown following:
\begin{equation}
  {{{\mathbf{X}^{1}}}_{*,j,i}}=\tau \underbrace{(\Gamma \overbrace{{{\mathbf{W}}_{*,j,c}}}^{ \mathit{Gen}-C} \Gamma \sigma \overbrace{({{{\tilde{\mathbf{W}}}}_{*,c,i}}{{U}_{*,c,i}}))}^{Pixel-C}}_{ \sum \subseteq \mathbbm{R}^{layer}}
\end{equation}
we focus on high cohesion and low coupling of channel features that mix the Pixel-$C$ with ${\mathbf{W}}_{*,j,c}$ and Gen-$C$ with ${{\mathbf{W}}}_{*,c,i}$.\par
For U-shape interpret module, the value of $i$ is halved with one layer of the encoder, and we employed a three-layer to extract more features for each pixel. Then we also have decoder operation with skip connection to mix spatial position with encoder features.
\begin{equation}
\left\{\begin{matrix}
\begin{aligned}
{{\mathbf{X}}^{e}}_{*,c,k} &=\tau \left ({{{\mathbf{W}}_{*,c,k}}}({{\mathbf{X}}^{1}}_{*,c,i})) \right),\text{    }k=\frac{i}{2}\text{ }  or \text{ } i*2 \\
{{\mathbf{X}}^{d}}_{*,c,*} &=\tau \left ({{\mathbf{X}}^{e}}_{*,c,*}+{{\mathbf{X}}^{{\tilde{d}}}}_{*,c,*} \right),\qquad\tilde{d}=t \\ 
\end{aligned}
\end{matrix}\right.
\end{equation}
where $\mathbf{W}_{*,c,k}$ scale the Pixel-$C$ dimension to embed deeper semantic information. ${{\mathbf{X}}^{d}}_{*,c,*}$ represents the result of skip connection with location information and it has same dimension with ${{\mathbf{X}}^{{\tilde{d}}}}_{*,c,*}$.
\begin{table*}[t]
\centering
\caption{Adopt OA, kappa, std and p-value as the metric. Classification results on the three datasets. The highest accuracy in each row is shown in bold. (\;\faAnchor\; denotes significance level is reached as \emph{p-value$<$0.05)} } 
\vspace{-0.3cm}
\label{table1}
\setlength{\tabcolsep}{1mm}{\renewcommand\arraystretch{1.5}\begin{tabular}{c|ccc|ccc|ccc}
\toprule
 &
  \multicolumn{3}{c|}{\textbf{PaviaU} } &
  \multicolumn{3}{c|}{\textbf{Houston 2013}} &
  \multicolumn{3}{c}{\textbf{Houston 2018}} 
  \\ \midrule
  \hline
Methods &
  
  
  OA (\%) &
  Kappa(\%) &
\emph{std,\;p-value }  &
  OA(\%) &

  Kappa(\%) &
\emph{std,\;p-value }  &
  OA(\%) &

  Kappa(\%) & 
 \emph{std,\;p-value } 
  \\ \midrule

\textbf{U-Net}\cite{Ronneberger2015UNetCN}       & 89.19 & 90.45.94 & \textbf{$\pm$\emph{0.0042,\; 5.2174e-4}} & 68.09 & 70.31 & \textbf{$\pm$\emph{0.0039,\; 9.0836e-4}}& 61.05& 59.96&\textbf{$\pm$\emph{0.0076,\; 3.5410e-4}} \\

\textbf{CAG(1}\%)\cite{Cai2020RemoteSI} & 92.13 & 90.05 & \textbf{$\pm$\emph{0.0194,\; 8.2395e-3}} & 83.42 & 85.21  & \textbf{$\pm$\emph{0.0129,\; 2.9732e-2}}& 70.28& 67.74&\textbf{$\pm$\emph{0.0114,\; 1.4363e-2}}\\

\begin{tabular}[c]{@{}c@{}}{\textbf{\footnotesize{HT-CNN-Att}}\cite{He2020HeterogeneousTL}}\end{tabular} &
  93.26 &
  91.51 &
  \textbf{$\pm$\emph{0.0172,\; 1.4910e-4}}&
  68.37 &
  68.73 & \textbf{$\pm$\emph{0.0191,\; 8.9163e-4}}& 68.37& 68.73&
\textbf{$\pm$\emph{0.0082,\; 3.4016e-3}}\\

\textbf{OTVCA}\cite{Rasti2017HyperspectralAL}         & 92.58 & 90.36 & \textbf{$\pm$\emph{0.0192,\; 3.3935e-2}} & 85.78 & 84.57& \textbf{$\pm$\emph{0.0141,\; 6.3231e-3}}& 72.63& 70.04&\textbf{$\pm$\emph{0.0122,\; 4.2511e-2}}\\

\textbf{CAGU}\cite{Lin2021ContextAwareAG}      & 89.98& 98.98& \textbf{$\pm$\emph{0.0067,\; 1.2123e-3}}& 99.96 & 23.90 & \textbf{$\pm$\emph{0.0081,\; 9.6180-4}}     & 80.62     & 79.06     & \textbf{$\pm$\emph{0.0093,\; 1.9281e-3}}     \\
\textbf{MLP-Mixer}\cite{Tolstikhin2021MLPMixerAA}         & \textbf{99.20 }&
  99.14 &
  \textbf{$\pm$\emph{0.0041,\; 7.2815e-4}} & 94.01 & 95.63 &\textbf{$\pm$\emph{0.0115,\; 3.4595e-3}} & 88.94 & 85.66 &\textbf{$\pm$\emph{0.0153,\; 9.9624e-3}} \\
  
\textbf{\ourM}           & \textbf{99.12}\textsuperscript{\faAnchor} & \textbf{99.24} &\textbf{$\pm$\emph{0.0021}} &

\textbf{96.01}\textsuperscript{\faAnchor}  & \textbf{95.58}& \textbf{$\pm$\emph{0.0117}}& \textbf{89.64}\textsuperscript{\faAnchor}& \textbf{88.64} &\textbf{$\pm$\emph{0.0108}}\\ \bottomrule
\end{tabular}}
 \vspace{0.1cm}
\end{table*}

\begin{table*}[t]
\centering
\caption{Adopt AA, std and p-value as the metric. Classification results on the three datasets. The highest accuracy in each row is shown in bold. (\;\faAnchor\; denotes significance level is reached as \emph{p-value$<$0.05)}} 
\vspace{-0.3cm}
\label{table2}
\setlength{\tabcolsep}{3.5mm}{\renewcommand\arraystretch{1.5}\begin{tabular}{c|cc|cc|cc}

\toprule
 &
  \multicolumn{2}{c|}{\textbf{PaviaU} } &
  \multicolumn{2}{c|}{\textbf{Houston 2013}} &
  \multicolumn{2}{c}{\textbf{Houston 2018}} 
   \\ \midrule
   \hline
Methods &
  
  
  AA (\%) &

\emph{std,\;p-value }  &
  AA(\%) &

\emph{std,\;p-value }  &
  AA(\%) &

 \emph{std,\;p-value } 
  \\ \midrule

\textbf{U-Net}\cite{Ronneberger2015UNetCN}        & 88.41 & \textbf{$\pm$\emph{0.0106,\; 9.7351e-4}}  & 68.09 & \textbf{$\pm$\emph{0.0251,\; 1.2635e-4}}&  60.72&\textbf{$\pm$\emph{0.0291,\; 7.1539e-4}} \\

\textbf{CAG(1}\%)\cite{Cai2020RemoteSI} & 86.34 & \textbf{$\pm$\emph{0.0137,\; 3.1846e-2}} & 85.06  & \textbf{$\pm$\emph{0.0094,\; 7.2649e-3}}& 67.39&\textbf{$\pm$\emph{0.0718,\; 6.2951e-3}}\\

\begin{tabular}[c]{@{}c@{}}{\textbf{\footnotesize{HT-CNN-Att}}\cite{He2020HeterogeneousTL}}\end{tabular} &
 
  86.34 &
  \textbf{$\pm$\emph{0.0620,\; 5.2940e-3}}&

  86.93 & \textbf{$\pm$\emph{0.0103,\; 2.1954e-2}}& 69.01&
\textbf{$\pm$\emph{0.0743,\; 2.1827e-3}}\\

\textbf{OTVCA}\cite{Rasti2017HyperspectralAL}        & 91.64 & \textbf{$\pm$\emph{0.0173,\; 2.8543e-2}} & 87.37& \textbf{$\pm$\emph{0.0860,\; 7.8496e-3}}&  69.48&\textbf{$\pm$\emph{0.0144,\; 9.3710e-2}}\\

\textbf{CAGU}\cite{Lin2021ContextAwareAG}     & 99.31& \textbf{$\pm$\emph{0.0013,\; 1.4618e-3}} & 95.92 & \textbf{$\pm$\emph{0.0094,\; 9.5386e-4}}          & 77.04    & \textbf{$\pm$\emph{0.0172,\; 2.5491e-3}}     \\
\textbf{MLP-Mixer}\cite{Tolstikhin2021MLPMixerAA}         & \textbf{99.31 }&
 
  \textbf{$\pm$\emph{0.0063,\; 9.0422e-3}}  & 94.48 &\textbf{$\pm$\emph{0.0179,\; 2.2106e-2}}  & 78.18 &\textbf{$\pm$\emph{0.0118,\; 5.5230e-3}} \\
  
\textbf{\ourM}           & \textbf{99.41}\textsuperscript{\faAnchor}  &\textbf{$\pm$\emph{0.0037}} &

\textbf{96.56}\textsuperscript{\faAnchor}  & \textbf{$\pm$\emph{0.0092}}& \textbf{83.65}\textsuperscript{\faAnchor} &\textbf{$\pm$\emph{0.0195}}\\ \bottomrule
\end{tabular}}
 \vspace{-0.5cm}
\end{table*}
In the UMLP block, we also employed stacking UMLP layers to obtain a deeper receptive field. 

%% file: section/3-experiments.tex
\section{EXPERIMENTS}
\subsection{Data description and details}
In this experiment, chose three widely used HSI datasets to evaluate the proposed methods' performance: Pavia University (PaviaU), Houston 2013 and Houston 2018. 
\begin{figure*}[htb]
    
    \includegraphics[width=1.0\textwidth]{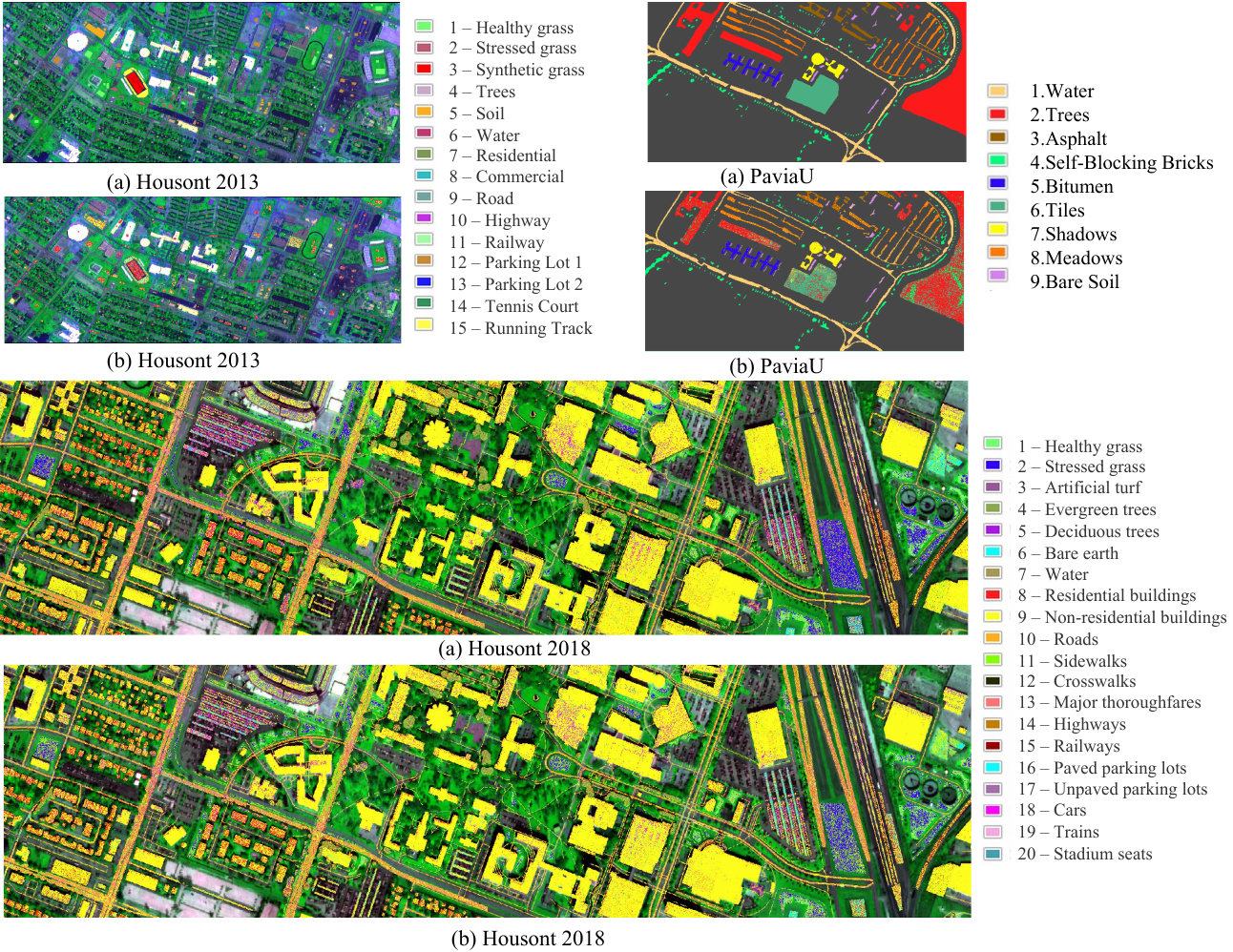}
    \vspace{-0.7cm}
    
    \caption{Classification maps for all datasets. (a) \ourM~Classification map. (b) HT-CNN-Attention Classification map.}
    \label{fig:experience}
    \vspace{-0.4cm}
\end{figure*}
The PaviaU is a hyperspectral image obtained by scanning the Italian city of Pavia in 2003.
%
%
Houston 2013 contains 144 spectral bands and covers a spatial region of 349 × 1905 pixels with a spatial resolution of 2.5m per pixel.
Houston 2018 covers 380-1050 nm spectral range with 48 bands at a 1-m GSD. For Houston 2018, the pixel size of the ground truth is 0.5.
Hence, we used the nearest neighbour algorithm to convert the pixel size to 1, corresponding to the training dataset.
We adopted random sampling to reasonably split the training dataset, validation dataset and test dataset.
All experiments are conducted on a Tesla V100 with 32GB RAM. 
The learning rate is 0.0002, the weight decay rate is 8e-7, and the learning rate decay function is LambdaLR.
\subsection{Result}
In this letter, we employed five metrics, namely OA, AA, Kappa, std and p-value, to evaluate the model's performance on the three well-known datasets. 
The Table~\ref{table1} and Table~\ref{table2} show that \ourM~ performance is superior for all comparison methods. 
Moreover, the visualization of the classification maps is shown in Fig.~\ref{fig:experience}.
\par
From Table~\ref{table1}, the overall accuracy of the \ourM~and MLP-Mixer is close, only a difference of 0.7\%, 2\% and 0.08\% in Houston 2018, Houston 2013 and paviaU, respectively. Furthermore, \ourM~also reached the level of significance. 
From Table~\ref{table2}, the average accuracy is always lower than the overall accuracy due to the number of sample pixels. \ourM~outperforms state-of-the-art methods across-the-board, improving 6.61\% on CAGU, 5.47\% on MLP-Mixer, 14.17\% on OTVCA, 14.64\% on HT-CNN-Attention in Houston 2018 dataset.
\par Analyzing these results, \ourM~ obtains a basic baseline (MSC module) through the operation of the rotating hybrid perceptron. Compared with those existing model with a certain bottleneck,such as HT-CNN-Attention, it perfectly extracts the hybrid representation of various features from different angles. Meanwhile, the U-shape module interprets the features from the MSC phase, and digs out the inherent logical relationship between the pixel channels. That is why \ourM~is superior compared with those existing model with a certain bottleneck. We conducted ablation experiments on the Houston 2018 data set, as shown in Table~\ref{ablation}. The results prove that our model is robust.
In order to ensure the robustness of the model, we calculated the std and p-value. The standard deviation of each dataset is less than 0.02 while reaching the significance level.
%

\begin{table}[]
\centering
\caption{Ablation of \ourM: MSC-2 block in \ourM~low-resolution. Experiment on Houston 2018 dataset.  (\;\faCheck\; denotes fixing and \;\faEyeSlash\; is opposite.} 
\label{ablation}
\setlength{\tabcolsep}{3.5mm}\renewcommand\arraystretch{1.5}\begin{tabular}{ccc|cc}
\hline
MSC-1 & MSC & UMLP & OA(\%)    & AA(\%)    \\ \hline
\faCheck     & \faCheck   & \faEyeSlash    & 87.41 & 82.38 \\
\faCheck     & \faEyeSlash   & \faCheck    & 88.29 & 82.76 \\ \hline
\end{tabular}
\vspace{-0.4cm}
\end{table}

%% file: section/4-conclusion.tex
\section{CONCLUSION}
We propose \ourM~in this letter, efficiency and effectiveness structure that the parameter scale is only 0.817M and outperforms state-of-the-art methods across-the-board.
Convolutional neural networks and attention-based structures can achieve outstanding performance, but neither of them is necessary.
Not surprising, we demonstrate that \ourM~is superior to state-of-the-art methods 
not only effectiveness but efficiency, which perfectly illustrates the extreme superiority in our approach.

%% file: section/5-Acknowledgement.tex
\vspace{-0.5cm}
\section{Acknowledgement }
We would like to thank the National Center for Airborne Laser Mapping and the Hyperspectral Image Analysis Laboratory at the University of Houston, Prof. Paolo Gamba from the Telecommunications and Remote Sensing Laboratory, Pavia University for acquiring and providing the data used in this study and thank for the IEEE GRSS Image Analysis and Data Fusion Technical Committee.